\documentclass[prd,showpacs,preprintnumbers,twocolumn,amsmath,nofootinbib,amssymb,floatfix]{revtex4}
\usepackage{graphicx,color,dcolumn,booktabs,bm}
\usepackage{epstopdf}
\usepackage{longtable,lscape}
\usepackage{txfonts}
\usepackage{overpic}
\usepackage{float}
\usepackage{amssymb}
\usepackage{amsmath}
\usepackage{epstopdf}
\usepackage{indentfirst}
\usepackage{feynmf}  
\usepackage{slashed}  
\usepackage{cases}
\usepackage{ulem}
\usepackage{color}
\usepackage{float}
\usepackage{multirow}
\usepackage{tikz}
\usepackage{epstopdf}
\usepackage{epsfig,dsfont,amssymb,amsmath,amsfonts,amsbsy,mathrsfs}
\usepackage{multirow}
\usepackage{graphicx}  
\usepackage{float}  
\usepackage{subfigure}  
\usepackage{array}
\usepackage{microtype}
\usepackage{pdfpages} 

\usepackage[colorlinks, citecolor=blue,anchorcolor=red,menucolor=red, linkcolor=red,filecolor=red,runcolor=red,urlcolor=blue,frenchlinks=true]{hyperref}
\makeatletter
\@addtoreset{equation}{section}
\makeatother

\newcommand{\nc}{\newcommand}
\nc{\tj}[1]{\textcolor{red}{Tianjin: #1}}

\usepackage{soul}
\allowdisplaybreaks
\begin{document}\title{ Construction of $a_4$ family}
\author{Ya-Rong Wang$^{1,2}$} 
 \author{Cheng-Qun Pang$^{1,4}$ }\email{xuehua45@163.com}
\author{Hao Chen$^{3}$ }\email{chenhao$\_$qhnu@outlook.com}
 \author{Xiao-Hai Liu$^{2}$ }\email{ xiaohai.liu@tju.edu.cn}
\affiliation{
$^1$ School of Physics and Optoelectronic Engineering, Ludong University, Yantai 264000, China\\
$^{2}$Center for Joint Quantum Studies and Department of Physics, School of Science, Tianjin University, Tianjin 300350, China\\
$^3$College of Physics and Electronic Information Engineering, Qinghai Normal University, Xining 810000, China\\
$^4$Lanzhou Center for Theoretical Physics, Key Laboratory of Quantum Theory and Applications of MoE, and Key Laboratory of Theoretical Physics of Gansu Province, Lanzhou University, Lanzhou, Gansu 730000, China
}
\date{\today}

\begin{abstract}

The COMPASS Collaboration recently reported the observation of a new resonance, $a_4(2610)$,  which has sparked our interest in studying the $a_4$ family with {$I^GJ^{PC}=1^{-}4^{++}$}. In this work, we investigate the mass spectra and Okubo-Zweig-Iizuka-allowed two-body strong decays of the $a_4$ family using the modified Godfrey-Isgur quark model and the quark-pair creation model. We also explore the possibility of identifying $a_4(2610)$ as a $4F$ or $2H$ state, {and our numerical results suggest that it could be a promising candidate for the $a_4(2H)$ state. In addition, we predict the masses and the widths of the $a_4(1H)$ and $a_4(3F)$ states.}

\end{abstract}
\maketitle

\section{introduction}\label{sec1}

Very recently, the COMPASS Collaboration announced a new resonance [{we denote it as $a_4(2610)$}], with a mass around 2.6 GeV in the $KK$ final state~\cite{Beckers2025}.
The corresponding mass and width were determined to be $2608 \pm 9^{+5}_{-38}$ MeV and $609 \pm 22^{+35}_{-311}$ MeV, respectively.
Meanwhile, we note that the PDG “Further States" also lists an $a_4(2255)$ meson \cite{ParticleDataGroup:2024cfk}, which has sparked our interest in investigating the internal structure of both $a_4(2255)$ and  $a_4(2610)$. Together with the ground state, the $a_4(1970)$ meson, establishing the $a_4$ meson family would be highly significant for completing the light meson family.

\par
The $a_4$ meson family  is characterized by {$I^GJ^{PC}=1^{-}4^{++}$}, which corresponds to spin $S=1$ and total angular momentum $J=4$. Due to the constraints from parity and spin-orbit coupling, the relative orbital angular momentum can take two possible values, $L = 3$ and $L = 5$, corresponding to the $F$-wave and $H$-wave, respectively.

In 1977, the Omega Group at CERN observed an $a_4$ state with a mass of $2030$ MeV and a width of $510\pm200$ MeV in a partial wave analysis (PWA) of $3\pi$ system production in the reaction $\pi^{-}p\to 3\pi n$ at 12 and 15 GeV/$c$~\cite{Corden:1977em}, which is the $a_4(1970)$ listed in PDG \cite{ParticleDataGroup:2024cfk}. 
$a_4(1970)$ was also found in the reactions {$\pi^{\pm} p \to K_s^0K^{\pm}p$}~\cite{Cleland:1982te}, $\pi^- p \to \eta \pi^0 n$~\cite{GAMS:1995rpx}, $\pi^- A \to \omega \pi^-\pi^0 A^*$~\cite{VES:1999olf}, $\pi^- p \to \eta' \pi^-p$~\cite{E852:2001ikk}, $\pi^- p \to \omega\pi^-\pi^0p$~\cite{E852:2004rfa}, and
$\pi^- P_b \to \omega\pi^-\pi^-\pi^+P_b'$~\cite{E852:2004rfa}.  Further evidence for this state was also  found in many other experiments~\cite{Baldi:1978ju,Anisovich:2001pn,Chung:2002pu,Uman:2006xb,COMPASS:2009xrl,COMPASS:2014vkj,COMPASS:2018uzl}. The latest observation of $a_4(1970)$ was reported by the COMPASS Collaboration,  {which  provides the most precise measurement of the resonance parameters $M=1952.2\pm1.8^{+3}_{-3.5}$ MeV and $\Gamma=324\pm4\pm6$ MeV~\cite{Beckers2025}}.

In 2001, Anisovich {\it{et al.}} reported evidence for $a_4(2255)$ in the reactions $p\bar{p} \to  \pi^0\eta, 3\pi^0, \text{and}~  \pi^0\eta^\prime$~\cite{Anisovich:2001pn}. {Uman {\it{et al.}}   confirmed the state $a_4(2255)$ use the data from the Fermilab E835 experiment} in the reaction $p\bar{p} \to \eta \eta\pi^0$~\cite{Uman:2006xb}.

$a_4(1970)$ is well established  as  the ground state of  the  $a_4$ family~\cite{Bramon:1980xm,Godfrey:1985xj,Ebert:2009ub,ParticleDataGroup:2024cfk}.
In our previous work,  $a_4(2255)$  was suggested to be the $2^3F_4$ state,   the first radial excitation of $a_4(1970)$~\cite{Pang:2015eha},  which is consistent with the conclusions of Refs.~\cite{Afonin:2007aa, Anisovich:2000kxa,Anisovich:2002us,Anisovich:2003tm, Masjuan:2012gc}.

Ebert {\it et al.} obtained a $1^3H_4$ state with a  mass of $M = 2234$ MeV, close to  that of $a_4(2255)$~\cite {Ebert:2009ub}. However, the PWA in Ref.~\cite{Anisovich:2001pn} showed that $a_4(2255)$ corresponds to a $^3F_4$ state. 

Now, with the discovery of $a_4(2610)$,  the $a_4$ family already contains three members: $a_4(1970)$, $a_4(2255)$, and $a_4(2610)$,   as listed in Table \ref{massinpdg}.
The study of internal structure  of $a_4(2610)$  and the construction of the $a_4$ family has become an interesting and useful issue. 

Mass spectral analysis and two-body strong decay are always used to probe the internal structure of mesons.
{Strong interactions in light meson systems cannot be calculated directly from first principles (except through Lattice Quantum Chromodynamics(QCD)), owing to the non-perturbative nature of QCD \cite{Prelovsek:2025gmd}. 
Consequently, phenomenological frameworks such as QCD sum rules, potential models \cite{Godfrey:1985xj,Ebert:2009ub}, and the $^3P_0$ model \cite{Micu:1968mk} are routinely employed to describe their spectra and decays.
In fact, both lattice calculations and QCD sum rules remain challenging, especially for higher excitations.  
} Phenomenological approaches are commonly adopted to explore their properties. 
For example, the Godfrey-Isgur (GI) potential model is often used to investigate meson spectra~\cite{Godfrey:1985xj,Godfrey:2016nwn},
While the GI model adequately reproduces the spectra of low-lying mesons, its extension to highly excited states requires the inclusion of color screening effects~\cite{Laermann:1986pu,Born:1989iv,Knechtli:2000df,Chao:1992et,Ding:1993uy}.  Song \textit{et al.} modified the GI  model (named as the MGI model) {by} taking into account the color screening effects, which has proven successful in describing charmed mesons~\cite{Song:2015nia}. Subsequently, this framework has been successfully applied to  doubly heavy flavor mesons, light mesons and baryons~\cite{Wang:2018rjg,Wang:2019mhs,Weng:2024roa}. 
For describing the nature of  higher excitations of the $a_4$ family, the MGI model is adopted   to study their property.
The {quark pair creation (QPC) model, also known as the $^3P_0$ model,} is an effective model to study the two-body strong {decays} of mesons~\cite{Micu:1968mk,Ackleh:1996yt}, which is well suited for this work. 
 
In our previous work, we adopted the MGI  model and {the} QPC model to predict the spectrum and two-body decay properties of {the} $5^{++}$ meson family, respectively~\cite{Pang:2019ttv}. In this work,  {these two models are  suited  to study} the  spectrum and two-body {decays} of {the} $a_4$ family. We hope that our effort will be helpful {in revealing} the internal structure of $a_4(2255)$ and $a_4(2610)$, and {in establishing the} $a_4$ meson family.

\par
The paper is organized as follows: In Sec. \ref{sec2}, we briefly {review}  the MGI model and the $^3P_0$ model.
In Sec. \ref{sec3}, {our numerical} results for  {the} $a_4$ family are presented. We first verify the assignment of the \sout{$a_4(2040)$}  {$a_4(1970)$} and the $a_4(2255)$. Then, we  study the newly observed $a_4(2610)$ and {provide a prediction of the $a_4(1H)$ and $a_4(3F)$}. Finally, a conclusion is given in Sec. \ref{sec4}.

 \begin{table}[htbp]
  \renewcommand{\arraystretch}{1.5}
  \centering
  \caption{Resonance parameters of the $a_4$ states. The units of the mass and width are MeV.}
  \label{massinpdg}
    \vspace{15pt}
  \begin{tabular}{lcc}
    \hline\hline
    State & Mass & Width \\ \hline
$a_4(1970)$  & $1967 \pm 16$  & $324^{+15}_{-18}$~\cite{ParticleDataGroup:2024cfk} \\
    \multirow{2}{*}{$a_4(2255)$}
                  & $2237 \pm 55$                         & $291 \pm 12$~\cite{Uman:2006xb}\\
                  & $2255 \pm 40$                         & $330^{+110}_{-50}$~\cite{Anisovich:2001pn} \\
    $a_4(2610)$   & $2608 \pm 9^{+5}_{-38}$              & $609\pm 22^{+35}_{-311}$~\cite{Beckers2025} \\
    \hline\hline
  \end{tabular}
\end{table}

\section{Models employed in this work} \label{sec2}

In this section, we introduce the MGI model and the QPC model employed in this work.
 
\subsection{The modified GI model}

\par
Based on the GI  model, which  {was} proposed by Godfrey and Isgur in 1985 {to describe} relativistic meson spectra with great success, especially for low-lying mesons~\cite{Godfrey:1985xj}, Song {\it{et al.}} proposed the MGI model~\cite{Song:2015nia}, in which {a} screened potential term was introduced to describe the charmed and charmed-strange meson's excited states {better}. Since then, the MGI model {has been} applied to study the light meson spectroscopy~\cite{Pang:2018gcn,Wang:2022xxi,Wang:2021abg,Feng:2022hwq,Wang:2024yvo,Wang:2024lba}, 
double heavy quarkonium~\cite{Wang:2018rjg,Wang:2019mhs}, and baryons~\cite{Weng:2024roa}. The Hamiltonian of the MGI model reads:
\begin{equation}\label{hh}
\tilde{H}=\sum_i{({m_i^2+\mathbf{p}_i^2})}^{1/2}+\tilde{V}^{\mathrm{eff}},
\end{equation}
where $m_i$ denotes the mass of the quark (or antiquark). The mass of the strange quark (or antiquark) $m_s=0.377$ GeV, the up and down quark (or antiquark) $m_{u(d)}=0.162$ GeV {are} adopted. {The effective potential} $\tilde{V}^{\mathrm{eff}}$ includes the following items:
\begin{equation}\label{V}
\tilde{V}^{\mathrm{eff}}=\tilde{G}_{12}+\tilde{V}^{\mathrm{cont}}+\tilde{V}^{\mathrm{tens}}+\tilde{V}^{\mathrm{so(v)}}+\tilde{S}_{12}(r)+\tilde{V}^{\mathrm{so(s)}},
\end{equation}
{where the individual terms are interpreted as follows: 

\begin{itemize}
  \item $\tilde{G}_{12}$ is referred to as the Coulomb term, which is understood to arise from one-gluon exchange.  
  \item $\tilde{V}^{\mathrm{cont}}$ is referred to as the contact term, which is introduced to account for the short-range interaction between quarks.  
  \item $\tilde{V}^{\mathrm{tens}}$ is denoted as the tensor term, which is generated by the spin--spin interaction from one-gluon exchange.  
  \item $\tilde{V}^{\mathrm{so(v)}}$ is referred to as the vector spin--orbit term.  
  \item $\tilde{S}_{12}(r)$ is denoted as the screened confinement term, which is introduced to reflect the color-screening effect.  
  \item $\tilde{V}^{\mathrm{so(s)}}$ is referred to as the scalar spin--orbit term.  
\end{itemize}

The explicit forms and detailed discussions of these potential terms are presented below.}

The spin-independent terms of the nonrelativistic  potential {are given by}
\begin{equation}
\tilde{G}(r)=-\sum_k\frac{4\alpha_k}{3r}\left[\frac{2}{\sqrt{\pi}}\int_0^{\gamma_{k}r} e^{-x^2}dx\right],
\end{equation}
with $\alpha_k=(0.25, 0.15, 0.2)$ and $\gamma_k=(1/2{\,\text{GeV}},\sqrt{10}/2{\,\text{GeV}},\sqrt{1000}/2{\,\text{GeV}})$ for $k=1,2,3$~\cite{Godfrey:1985xj}, 
and
\begin{equation}
{S}(r)=\frac{b(1-e^{-\mu r})}{\mu}+c,
\end{equation}
where $\mu = 0.0779$ {GeV} is {adopted as} the screening parameter from our previous work~\cite{Wang:2024lba}. This parameter characterizes the strength of the color screening effect, which is absent in the GI model, and thus represents an improvement of the MGI model.
{The confining parameter is taken as $b=0.222$ GeV$^2$, while the vacuum constant is $c=-0.228$ GeV~\cite{Wang:2024lba}.}

There are two {methods} to characterize the effective potential $\tilde{V}^{\mathrm{eff}}$ {when relativistic effects in meson systems are taken into account. The first {method} introduces a smearing function, defined as follows:}
\begin{equation}
\label{smearing}
\rho_{ij} \left(\mathbf{r}-\mathbf{r'}\right)=\frac{\sigma_{ij}^3}{\pi ^{3/2}}e^{-\sigma_{ij}^2\left(\mathbf{r}-\mathbf{r'}\right)^2},
\end{equation}
with
\begin{align}
\label{smearing_sigma}
   \sigma_{ij}^2=\sigma_0^2\Bigg[\frac{1}{2}+\frac{1}{2}\left(\frac{4m_im_j}{(m_i+m_j)^2}\right)^4\Bigg]+
  s^2\left(\frac{2m_im_j}{m_i+m_j}\right)^2,
\end{align}
where $\sigma_0=1.791$ GeV is a universal parameter and $s=0.711$. {The values of these two parameters, listed in Table~\ref{MGI},} are taken from Ref.~\cite{Wang:2024lba}.
The Coulomb term $\tilde{G}_{12}(r)$ is {then written} as
\begin{equation}
\begin{split}
\tilde{G}_{ij}(r)=&\int {\rm{d}}^3{\bf r}^\prime \rho_{ij}({\bf r}-{\bf r}^\prime)G(r^\prime)
=\sum\limits_k-\frac{4\alpha_k }{3r}{\rm erf}(\tau_{kij}r),
\end{split}
\end{equation}
where
\begin{equation}
\tau_{kij}=\frac{1}{\sqrt{\frac{1}{\sigma_{ij}^2}+\frac{1}{\gamma_k^2}}}.
\end{equation}
The {confinement} potential $\tilde{S}_{12}(r)$ can be expressed as
\begin{eqnarray}
\tilde{S}_{12}(r)&=& \int {\rm{d}}^3 {\bf r}^\prime
\rho_{12} ({\bf r}-{\bf r}^\prime)S(r^\prime)\nonumber\\
&=& \frac{b}{\mu r}\Bigg[r+e^{\frac{\mu^2}{4 \sigma^2}+\mu r}\frac{\mu+2r\sigma^2}{2\sigma^2}\Bigg(\frac{1}{\sqrt{\pi}}
\int_0^{\frac{\mu+2r\sigma^2}{2\sigma}}e^{-x^2}{\rm{d}}x-\frac{1}{2}\Bigg) \nonumber\\
&&-e^{\frac{\mu^2}{4 \sigma^2}-\mu r}\frac{\mu-2r\sigma^2}{2\sigma^2}\Bigg(\frac{1}{\sqrt{\pi}}
\int_0^{\frac{\mu-2r\sigma^2}{2\sigma}}e^{-x^2}{\rm{d}}x-\frac{1}{2}\Bigg)\Bigg]  \nonumber \\
&&+c. \nonumber\label{Eq:pot}
\end{eqnarray}
{The second method introduces} the momentum-dependent factors
\begin{equation}
\tilde{G}_{12}(r)\to \tilde{G}_{12}=\left(1+\frac{\mathbf{p}^2}{E_1E_2}\right)^{1/2}\tilde{G}_{12}(r)\left(1 +\frac{\mathbf{p}^2}{E_1E_2}\right)^{1/2}.
\end{equation}
The semirelativistic  {corrections} of the spin-dependent terms are  written as
\begin{equation}
\label{vsoij}
  \tilde{V}^i_{\alpha \beta}(r)\to\tilde{V}^i_{\alpha \beta}= \left(\frac{m_\alpha m_\beta}{E_\alpha E_\beta}\right)^{1/2+\epsilon_i} \tilde{V}^i_{\alpha \beta}(r)\left(\frac{m_\alpha m_\beta}{E_\alpha E_\beta}\right)^{1/2+\epsilon_i},
\end{equation}
where $\tilde{V}^i_{\alpha \beta}(r)$  {denote} the contact term, the tensor term, the vector, and the scalar spin-orbit terms. The parameters $\epsilon_i=\epsilon_c$, $\epsilon_t$, $\epsilon_{\rm so(v)}$, and $\epsilon_{\rm so(s)}$ represent the relativistic corrections to $\tilde{V}^{\mathrm{cont}}$, $\tilde{V}^{\mathrm{tens}}$, $\tilde{V}^{\mathrm{so(v)}}$, and $\tilde{V}^{\mathrm{so(s)}}$, respectively~\cite{Wang:2021abg}. {Then the explicit forms of the spin-dependent potentials are}
\begin{equation}\label{Vcont}
\begin{split}
\tilde{V}^{\mathrm{cont}}=\frac{2{\bf S}_1\cdot{\bf S}_2}{3m_1m_2}\nabla^2\tilde{G}_{12}^c,
\end{split}
\end{equation}
\begin{equation}\label{Vtens}
\begin{split}
\tilde{V}^{\mathrm{tens}}=&-\left(\frac{3{\bf S}_1\cdot{\bf r}{\bf S}_2\cdot{\bf r}/r^2-{\bf S}_1\cdot{\bf S}_2}{3m_1m_2}\right)\left(\frac{\partial^2}{\partial r^2}-\frac{1}{r}{\frac{\partial}{\partial r}}\right)\tilde{G}_{12}^t,
\end{split}
\end{equation}
\begin{equation}\label{Vsov}
\begin{split}
\tilde{V}^{\mathrm{so(v)}}=&\frac{{\bf S}_1\cdot {\bf L}}{2m_1^2}\frac{1}{r}\frac{\partial\tilde{G}_{11}^{\rm so(v)}}{\partial r}+\frac{{\bf S}_2\cdot {\bf L}}{2m_2^2}\frac{1}{r}\frac{\partial\tilde{G}_{22}^{\rm so(v)}}{\partial r}
\\
 &
+\frac{({\bf S}_1+{\bf S}_2)\cdot {\bf L}}{m_1m_2}\frac{1}{r}\frac{\partial\tilde{G}_{12}^{\rm so(v)}}{\partial r},\\
\end{split}
\end{equation}
\begin{equation}\label{Vsos}
\begin{split}
\tilde{V}^{\mathrm{so(s)}}=&-\frac{{\bf S}_1\cdot {\bf L}}{2m_1^2}\frac{1}{r}\frac{\partial\tilde{S}_{11}^{\rm so(s)}}{\partial r}-\frac{{\bf S}_2\cdot {\bf L}}{2m_2^2}\frac{1}{r}\frac{\partial\tilde{S}_{22}^{\rm so(s)}}{\partial r}.\\
\end{split}
\end{equation}
\begin{table}[htbp]
\renewcommand{\arraystretch}{1.5}
\caption{Parameters of the MGI model~\cite{Wang:2024lba}. $m_{u(d)}$ and $m_s$ are the masses of the $u(d)$ and $s$ quarks (or antiquarks), $\mu$ is the screening parameter, $b$ is the confining  parameter, {and} $c$ is the vacuum constant. $\sigma_0$ is {the} universal parameter in Eq.~(\ref{smearing_sigma}), {while} $s$ is a parameter related to heavy quarkonium masses. $\epsilon_c$, $\epsilon_t$, $\epsilon_{\rm so(v)}$, and $\epsilon_{\rm so(s)}$ represent the relativistic corrections to the potential terms $\tilde{V}^{\mathrm{cont}}$, $\tilde{V}^{\mathrm{tens}}$, $\tilde{V}^{\mathrm{so(v)}}$, and $\tilde{V}^{\mathrm{so(s)}}$, respectively.\label{MGI}}
\begin{center}
\begin{tabular}{cccc}
\hline\hline
Parameter &  value &Parameter &  value  \\
 \midrule[0.7pt]          
$m_{u(d)}$(GeV)    &0.162    &{$s$ }          &{0.711}\\
$m_s$ (GeV)       &0.377    &$\mu$ (GeV)          &0.0779 \\
$b$ (GeV$^2$)     &0.222    &$c$ (GeV)            &$-0.228$\\
$\epsilon_c$      &-0.137   &$\epsilon_{so(v)}$     &0.0550\\
$\epsilon_{so(s)}$  &0.366    &$\epsilon_t$         &0.493\\
 {$\sigma_0$ (GeV)}   &{1.791}  &\dots   &\dots\\
  \hline\hline
\end{tabular}
\end{center}
\end{table}

We use the simple harmonic oscillator (SHO) basis to solve the spectrum of the light mesons. {The SHO wave functions are} given by
\begin{equation}
 \begin{split}
\psi_{nLM_L}^{SHO}(\mathbf{r})=R_{nL}^{SHO}(r, \beta)Y_{LM_L}(\Omega_r),\\
	\psi_{nLM_L}^{SHO}(\mathbf{p})=R_{nL}^{SHO}(p, \beta)Y_{LM_L}(\Omega_p),
\end{split}
\end{equation}
with
\begin{align} \label{1.3}
R_{nL}^{SHO}(r,\beta)=N_{nL}\beta^{3/2}(\beta r)^{L}
e^{\frac{-r^2 \beta^2}{2}} L_{n-1}^{L+1/2}(\beta^2r^2)
,\\
R_{nL}^{SHO}(p,\beta)=\frac{(-1)^{(n-1)}(-i)^L}{\beta^{3/2}}N_{nL}e^{-\frac{p^2}{2\beta^2}}{(\frac{p}{\beta})}^{L} \times L_{n-1}^{L+1/2}(\frac{p^2}{ \beta ^2}),
\end{align}
{and normalization factor}
\begin{align} \label{1.4}
N_{nL}=\sqrt{\frac{2{(n-1)}!}{\Gamma(n+L+1/2)}},
\end{align}
where $Y_{LM_L}(\Omega)$ {denotes the} spherical harmonic function, $L_{n-1}^{L+1/2}(x)$  the associated Laguerre polynomial and $\Gamma(n+L+1/2)$  the gamma function.
And 
\begin{align} \label{expand}
R_{nL}(r)=\sum_{n=1}^{n_{\rm max}}C_{n}{R}_{nL}^{\rm SHO}(r,\beta),\\
R_{nL}(p)=\sum_{n=1}^{n_{\rm max}}C_{n}{R}_{nL}^{\rm SHO}(p,\beta)
\end{align}
{are} the spatial wave {functions} of the mesons, where $C_{n}$ are the expansion coefficients, 
which can be derived through the process of diagonalizing the Hamiltonian (\ref{hh}). 
Then the SHO wave function depends only on a single parameter $\beta$, {which is determined by minimizing the eigenvalue $E_{nL}$, i.e., $\partial E_{nL}/\partial \beta_i=0$ and $\partial^2 E_{nL}/\partial \beta_i^2>0$, where $i$ labels different light mesons, and $n_{max}=21$ is adopted in the present calculation.} 
The spatial wave {functions} of the {mesons} obtained {with the} MGI model are {then} applied to the calculation of strong decay processes. 

\begin{table*}[htb] 
 \renewcommand{\arraystretch}{1.5}
\centering
{
\caption
{The mass spectra of the fitted light meson states. ``Expe." refers to the experimental value, with the units in MeV. The experimental values and most of the errors are taken from the PDG \cite{ParticleDataGroup:2024cfk}. \label{fit}}
\[\begin{array}{ccccccccc}
\hline
\hline
\text{State} & \text{Fit value} &\text{Expe. value}& \text{Fit error}&\text{State} & \text{Fit value} &\text{Expe. value}& \text{Fit error}\\
\hline
 \text{$\pi $(1S)} & 145.3 & 139.6 & 1 & \text{$\pi $(2S)} & 1277  & 1300 & 100 \\
 \text{$\pi $(3S)} & 1757 & 1812 & 12 & b_1\text{(1P)} & 1219 & 1230 & 3.2 \\
 h_1^\prime\text{(1P)}  & 1479 & 1417 & 8 & b_1\text{(3P)} & 2048 & 1960 & 30 \\
 b_1\text{(4P)} & 2320 & 2240 & 35 & \text{$\rho $(1S)} & 774.2 & 775.3 & 0.25 \\
 \text{$\rho $(2S)} & 1400 & 1465 & 25 & \text{$\rho $(3S)} & 1853 & 1900 & 30 \\
 \text{$\rho $(4S)} & 2170 & 2265 & 40 & a_0\text{(1P)} & 1142 & 1474 & 19 \\
 a_1\text{(1P)} & 1210 & 1230 & 40 & a_1\text{(2P)} & 1717 & 1647 & 22 \\
 a_2\text{(1P)} & 1317 & 1318 & 0.5 & a_2\text{(2P)} & 1739 & 1732 & 16 \\
 \pi _2\text{(1D)} & 1651 & 1672 & 3 & \pi _2\text{(2D)} & 2000 & 2005 & 15 \\
 \pi _2\text{(3D)} & 2273 & 2285 & 32 & b_3\text{(1F)} & 1959 & 2032 & 12 \\
 b_3\text{(2F)} & 2235 & 2245 & 50 & \rho _2\text{(1D)} & 1644 & 1940 & 40 \\
 \rho _2\text{(2D)} & 2001 & 2225 & 35 & \text{$\rho $(1D)} & 1612 & 1720 & 20 \\
 \text{$\rho $(2D)} & 1991 & 2000 & 30 & \text{$\rho $(3D)} & 2266 & 2265 & 40 \\
 \rho _3\text{(1D)} & 1684 & 1689 & 2.1 & \rho _3\text{(2D)} & 2015 & 1982 & 14 \\
 \rho _5\text{(1G)} & 2212 & 2330 & 35 & a_4\text{(1F)} & 1973 & 1996 & 10 \\
 a_4\text{(2F)} & 2243 & 2237 & 5 & a_6\text{(1H)} & 2415 & 2450 & 130 \\
 \phi _3\text{(1D)} & 1894 & 1854 & 7 & \text{$\phi $(1S)} & 1022 & 1020 & 1 \\
 {K(1S)} & 495.2 & 497.7 & 0.4 & {K(2S)} & 1449 & 1460 & 20 \\
 K^*\text{(1S)} & 911.5 & 895.8 & 0.8 & K^*\text{(2S)} & 1541 & 1414 & 15 \\
 K_0{}^*\text{(1P)} & 1284 & 1425 & 50 & K_2{}^*\text{(1P)} & 1440 & 1432 & 1.3 \\
 K_1{}^*\text{(1D)} & 1735 & 1717 & 27 & K_3{}^*\text{(1D)} & 1794 & 1776 & 7 \\
 K_4{}^*\text{(1F)} & 2075 & 2045 & 9 & K_5{}^*\text{(1G)} & 2309 & 2382 & 24 \\
 \hline
 \chi^2=41\\
 \hline
 \hline
\end{array}\]}
\end{table*}
{
Finally, it is important to note that the MGI model provides a global fit to the light meson spectrum with a $\chi^2$ value of about 40, as shown in  Table \ref{fit}. 
The 11 parameters used in the model are not set arbitrarily, but are determined by fitting to the experimental data.
Although this $\chi^2$ value is not particularly small in the context of data fitting, it remains reasonable within the framework of the potential models for meson spectroscopy. 
If we use the mean relative error of 1.3\% for the 44 experimental mass values, according to $\chi^2 = \left(\frac{M_{\text{Expe.}} - M_{\text{theory}}}{\text{error}}\right)^2$, one can obtain that the theoretical mass $M_{\text{theory}} \approx M_{\text{Expe.}} \times (1\pm 0.1)$, indicating a relative error of about 10\% in the fitting results.
Therefore, the internal structure of $a_4(2610)$ cannot be determined solely from its mass spectrum. Further investigation into its strong decay properties is required, which will be facilitated by the $^3P_0$ model introduced in the following section.
}

\subsection{The QPC model}
The QPC model, {also known as} $^3P_0$ model, was {first} proposed by Micu~\cite{Micu:1968mk} and further developed by the Orsay group~\cite{LeYaouanc:1972ae, LeYaouanc:1973xz, LeYaouanc:1974mr, LeYaouanc:1977gm, LeYaouanc:1977ux}. 
This model {has been} widely applied to the {calculation of} OZI-allowed two-body strong decays of mesons~\cite{vanBeveren:1982qb, Titov:1995si, Ackleh:1996yt, Blundell:1996as, Bonnaz:2001aj, Zhou:2004mw, Lu:2006ry, Zhang:2006yj, Luo:2009wu, Sun:2009tg, Liu:2009fe, Sun:2010pg, Rijken:2010zza, Ye:2012gu, Wang:2012wa, He:2013ttg, Sun:2013qca,  Pang:2018gcn, Wang:2022juf, Wang:2022xxi, Li:2022khh, Li:2022bre, Wang:2020due, Pang:2017dlw, Wang:2024yvo, Wang:2021abg, feng:2021igh, Feng:2022hwq}.
{{Recent studies~\cite{Bruschini:2025paj,Alkofer:2023syz} further provide theoretical support for the rationality of this model.}}
In this model, the {transition operator} $\mathcal{T}$ describes {the creation of} a quark-antiquark pair (denoted by indices 3 and 4) from {the} vacuum {with quantum numbers}
$^{2S+1}L_J=^{3}P_0$, and can be written as 
{\begin{align}\label{gamma}
\mathcal{T} = & -3\gamma \sum_{m}\langle 1m;1~-m|00\rangle\int d \mathbf{p}_3d\mathbf{p}_4\delta ^3 (\mathbf{p}_3+\mathbf{p}_4) \nonumber \\
 & ~\times \mathcal{Y}_{1m}\left(\frac{\textbf{p}_3-\mathbf{p}_4}{2}\right)\chi _{1,-m}^{34}\phi _{0}^{34}
\left(\omega_{0}^{34}\right)_{ij}b_{3i}^{\dag}(\mathbf{p}_3)d_{4j}^{\dag}(\mathbf{p}_4).
\end{align}
Here, the parameter $\gamma$ in QPC model {represents} the strength of $q\bar{q}$ {pair} creation from the vacuum, {and in this work }the value $\gamma  = 10.16$ {is adopted}~\cite{Wang:2024lba}. $\mathcal{Y}_l^m(\bf{p})\equiv$ $p^lY_l^m(\theta_p,\phi_p)$ {denotes} a solid harmonic. {The symbols} $\chi$, $\phi$, and $\omega$ denote the spin, flavor, and color wave functions, respectively. $\mathbf{p}_3$ and $\mathbf{p}_4$ {are} the three-momenta of the  {quark–antiquark pair created from the} vacuum{, while $i$ and $j$ are their color indices.} {With this transition operator, the decay amplitudes of mesons can be systematically calculated within the QPC framework.} The amplitude $\mathcal{M}^{{M}_{J_{\mathrm{A}}}M_{J_{\mathrm{B}}}M_{J_{\mathrm{C}}}}$  {is defined as}
\begin{equation}
\langle {\mathrm{BC}}|\mathcal{T}|{\mathrm{A}} \rangle = \delta ^3({\mathbf{P}_{\mathrm{B}}+\mathbf{P}_{\mathrm{C}})}\mathcal{M}^{{M}_{J_{\mathrm{A}}}M_{J_{\mathrm{B}}}M_{J_{\mathrm{C}}}},
\end{equation}
where $\mathbf{P}_{\mathrm{B}}$  and $\mathbf{P}_{\mathrm{C}}$ are the three-momenta of mesons $\mathrm{B}$ and $\mathrm{C}$ in the rest frame of the meson $\mathrm{A}$, {and} {${M}_{J_{X}}$ (with $X=\mathrm{A},\mathrm{B},\mathrm{C}$) denotes the magnetic quantum number of the corresponding meson.} Finally, the general form of the decay width can be expressed as
\begin{eqnarray}
\Gamma_{{\mathrm{A}}\to {\mathrm{BC}}}&=&\frac{\pi}{4} \frac{|\mathbf{P}|}{m_{\mathrm{A}}^2}\sum_{J,L}|\mathcal{M}^{JL}(\mathbf{P})|^2,
\end{eqnarray}
where $m_{\mathrm{A}}$ is the mass of the initial {meson} $\mathrm{A}$, {$\mathbf{J}=\mathbf{J}_{\mathrm{B}}+\mathbf{J}_{\mathrm{C}}$}, $L$ is the relative orbital angular momentum between {mssons $\mathrm{B}$ and $\mathrm{C}$}, and
$\mathbf{P}=\mathbf{P}_{\mathrm{B}}$. {The partial-wave amplitude $\mathcal{M}^{JL}(\mathbf{P})$ is related to the amplitude $\mathcal{M}^{{M}_{J_{\mathrm{A}}}M_{J_{\mathrm{B}}}M_{J_{\mathrm{C}}}}$ via the Jacob–Wick formula}~\cite{Jacob:1959at}}
\begin{equation}
\begin{aligned}
\mathcal{M}^{JL}(\mathbf{P}) = &\frac{\sqrt{4\pi(2L+1)}}{2J_{\mathrm{A}}+1}\sum_{M_{J_{\mathrm{B}}}M_{J_{\mathrm{C}}}}\langle L0;JM_{J_{\mathrm{A}}}|J_{\mathrm{A}}M_{J_{\mathrm{A}}}\rangle \\
    &\times \langle J_{\mathrm{B}}M_{J_{\mathrm{B}}};J_{\mathrm{C}}M_{J_{\mathrm{C}}}|{J_{\mathrm{A}}}M_{J_{\mathrm{A}}}\rangle \mathcal{M}^{M_{J_{{\mathrm{A}}}}M_{J_{\mathrm{B}}}M_{J_{\mathrm{C}}}},
    \end{aligned}	
\end{equation}
in which
\begin{equation}
\begin{aligned}
 &\mathcal{M}^{M_{J_{\mathrm{A}}} M_{J_{\mathrm{B}}} M_{J_{\mathrm{C}}}}\\
 &=\gamma \sqrt{8E_AE_BE_C}\sum_{\substack{M_{L_{\mathrm{A}}}, M_{S_{\mathrm{A}}}, M_{L_{\mathrm{B}}},\\ M_{S_{\mathrm{B}}}
M_{L_{\mathrm{C}}, M_S}, m}}\left\langle L_{\mathrm{A}} M_{L_{\mathrm{A}}} S_{\mathrm{A}} M_{S_{\mathrm{A}}} \mid J_{\mathrm{A}} M_{J_{\mathrm{A}}}\right\rangle \\
& \times\left\langle L_{\mathrm{B}} M_{L_{\mathrm{B}}} S_{\mathrm{B}} M_{S_{\mathrm{B}}} \mid J_{\mathrm{B}} M_{J_{\mathrm{B}}}\right\rangle\left\langle L_{\mathrm{C}} M_{L_{\mathrm{C}}} S_{\mathrm{C}} M_{S_{\mathrm{C}}} \mid J_{\mathrm{C}} M_{J_{\mathrm{C}}}\right\rangle \\
& \times\langle 1 m 1-m \mid 00\rangle\left\langle\chi_{S_{\mathrm{B}} M_{S_{\mathrm{B}}}}^{14} \chi_{S_{\mathrm{C}} M_{S_{\mathrm{C}}}}^{32} \mid \chi_{S_{\mathrm{A}} M_{S_{\mathrm{A}}}}^{12} \chi_{1-m}^{34}\right\rangle \\
& \times\left[\left\langle\phi_{\mathrm{B}}^{14} \phi_{\mathrm{C}}^{32} \mid \phi_{\mathrm{A}}^{12} \phi_0^{34}\right\rangle I\left(\mathbf{P}, m_1, m_2, m_3\right)\right.+ \\
& \left.(-1)^{1+S_{\mathrm{A}}+S_{\mathrm{B}}+S_{\mathrm{C}}+L_{\mathrm{C}}}\left\langle\phi_{\mathrm{B}}^{32} \phi_{\mathrm{C}}^{14} \mid \phi_{\mathrm{A}}^{12} \phi_0^{34}\right\rangle I\left(-\mathbf{P}, m_2, m_1, m_3\right)\right], \\
\end{aligned}
\end{equation}
with {the overlap integral}
\begin{equation}
\begin{aligned}
&I\left(\mathbf{P}, m_1, m_2, m_3\right)\\
&=\int d^3 \mathbf{k} \psi_{\mathrm{B}}^*(\mathbf{k}+U \mathbf{P}) \psi_{\mathrm{C}}^*(\mathbf{k}+V\mathbf{P} ) \psi_{\mathrm{A}}(\mathbf{k}-\mathbf{P}) \mathcal{Y}_1^m(\mathbf{k}),
\end{aligned}
\end{equation}
{where} 
\begin{equation*}
\begin{aligned}
U=\frac{m_3}{m_1+m_3}, V=\frac{m_3}{m_2+m_3}{.}
\end{aligned}
\end{equation*}
{Here,} $m_1$ and $m_2$ are the masses of the quark and the antiquark in meson {$\mathrm{A}$}, respectively. In this work, $m_1=0.377$ GeV, $m_2=0.162$ GeV for $s\bar n$ (or $n\bar s$), $m_1=m_2=0.162$ GeV for $n\bar n$, and $m_1=m_2=0.377$ GeV for $s\bar s$. The mass of the created quark (or antiquark) from the vacuum is denoted as $m_3$, and is taken as 0.162 GeV for $n\bar{n}$ and 0.377 GeV for $s\bar{s}$.

The spatial wave functions of mesons obtained using the MGI model are {employed in the calculation of} the strong decays of the $a_4$ family {within the} QPC model as we {mentioned} previously.

{For} the final states, the mixing scheme of strange mesons {with natural parity ($L=J$)} can be expressed as
\begin{equation}\label{anglek1}
\left( \begin{matrix}
	|K(nL)\rangle \\
	|K^\prime( nL)\rangle \\
\end{matrix}\right) =
\left( \begin{matrix}
	\textrm{$\cos\theta_{nL}$} & \textrm{$\sin\theta_{nL}$} \\
	\textrm{$-\sin\theta_{nL}$} & \textrm{$\cos\theta_{nL}$} \\
\end{matrix}\right)
\left( \begin{matrix}
	|K(n^1L_L)\rangle \\
	|K(n^3L_L)\rangle \\
\end{matrix}\right),
\end{equation}
where $\theta_{nL}$ {denotes} as the mixing angle between the $K(n^1L_L)$ and $K(n^3L_L)$ states. In this work, the masses of ${K(nL)}$ and ${K^\prime (nL)}$ are calculated {using} the MGI model~\cite{Pang:2025esm}.  
{The mixing angle is taken as} $\theta_{1P}=-34^\circ$~\cite{Cheng:2013cwa}. {For other cases, the mixing angle is given by} $\theta_{nL}=-\text{arctan}(\sqrt{\frac{L}{L+1}})$~\cite{Asghar:2019qjl}. 

The flavor wave functions of isoscalar mesons {can be expressed in the mixing form}
\begin{equation}\label{mixingns}
\left( \begin{matrix}
	X \\
	X^\prime \\
\end{matrix}\right) =
\left( \begin{matrix}
	\textrm{$\cos\phi_{x}$} & \textrm{$\sin\phi_{x}$} \\
	\textrm{$-\sin\phi_{x}$} & \textrm{$\cos\phi_{x}$} \\
\end{matrix}\right)
\left( \begin{matrix}
	|n\bar{n}\rangle \\
	|s\bar{s}\rangle \\
\end{matrix}\right),
\end{equation}
where $X$ and $X^\prime$ {denote} two isoscalar mesons (such as $\eta$ and $\eta^\prime$), $\phi_x$ is the mixing angle in the {quark-flavor} scheme, and {the light nonstrange component is defined as} $n\bar{n}=(u\bar{u}+d\bar{d})/\sqrt{2}$. The {flavor mixing} information of the isoscalar mesons {used} in this {work is} adopted from Ref.~\cite{Pang:2025esm}.

\section{Numerical results and phenomenological analysis}\label{sec3}
The spectrum of {the} $a_4$ family {is calculated using the MGI model and listed in Table}~\ref{mass}. 
The OZI-allowed two-body strong decay properties of {the} $a_4$ family {are presented} in Tables \ref{1F2F}-\ref{1H3F}. {We now turn to a phenomenological analysis of the spectrum and the decay information of the $a_4$ family.}

\subsection{Verification of the {assignment of} the \sout{$a_4(2040)$} {$a_4(1970)$} and the $a_4(2255)$}
As the ground state of the $a_4$ meson family, $a_4(1970)$ has been well established both theoretically and experimentally~\cite{Corden:1977em, Baldi:1978ju, Cleland:1982te, VES:1999olf, Anisovich:2001pn, Chung:2002pu, Bramon:1980xm, Godfrey:1985xj, Ebert:2009ub, ParticleDataGroup:2024cfk}.
\par
In 1978, M.~J.~Corden \textit{et al.} {collected} data on the charge-exchange reaction $\pi^-p \to 3\pi n$ at beam momenta of 12 and 15 GeV/$c$ using the CERN Omega Multiparticle Spectrometer. {A} natural spin-parity enhancement {was observed} at a mass of about 2 GeV/$c^2$ with $J^P=4^+$ preferred, which {was denoted} $A_2^*(2030)$ ($a_4(1970)$ in this work)~\cite{Corden:1977em}. 
In the same year, R.~Baldi \textit{et al.} {reported} the observation of {an} $I^GJ^P=1^-4^+$ ($a_4(1970)$ in this work) state in the reaction $\pi^-p \to K_S^0K^-p{)}$ at 10 GeV/$c$, measured with a nonmagnetic spectrometer at the CERN proton synchrotron (PS)~\cite{Baldi:1978ju}. {Evidence for the} $a_4(1970)$ meson  {was obtained} at  {a mass of about} $1900$ MeV with a width of about 200 MeV~\cite{Baldi:1978ju}. 
In 1982, W.~E.~Cleland \textit{et al.}  analysed the reaction $\pi^{\pm}p \to K_S^0K^{\pm}p$ at 30 and 50 GeV/$c$~\cite{Cleland:1982te}. This analysis confirmed the {spin-4} $A_2(2040)$ ($a_4(1970)$ in this work) state~\cite{Cleland:1982te}. {A signal with} nine standard {deviations}  in  both beam polarities at 50 GeV/$c$ {yielded} a $K\bar{K}$ cross section of {$0.50 \pm 0.09$ $\mu$b} for this resonance, {with mass and width determined as} $M=2040 \pm 30$ MeV and $\Gamma=380 \pm 150$ MeV~\cite{Cleland:1982te}.
By 1999, $a_4(1970)$ was once again experimentally observed in the {reaction} $\pi^-A \to \omega\pi^-\pi^0A^*$~\cite{VES:1999olf}.
Since {then,}  a considerable amount of experimental data on $a_4(1970)$ has emerged. In 2001, A.~V.~Anisovich \textit{et al.} reported a combined analysis of $3\pi^0$, $\pi^0\eta$ and $\pi^0\eta^\prime$ data in the mass range {1960–2410} MeV and found $a_4(1970)$ {again} with mass and width of $2005^{+25}_{-45}$ MeV and $180\pm30$ MeV~\cite{Anisovich:2001pn}. 
Subsequently, $a_4(1970)$ was verified in  reactions $\pi^-p \to \eta^\prime \pi^- p$~\cite{E852:2001ikk}, $\pi^-p \to 3 \pi p$~\cite{Chung:2002pu}, $\pi^-p \to \omega\pi^-\pi^0p$~\cite{E852:2004rfa}, {and}
$\bar{p}p \to \eta\eta\pi^0$~\cite{Uman:2006xb}.
 In 2010, the COMPASS experiment at the CERN SPS studied the diffractive dissociation of negative pions into the $\pi^-\pi^-\pi^+$ final state using a 190 GeV/$c$ pion beam {on} a lead target{, and clearly confirmed the} $a_4(1970)$ with {a mass of $1885 \pm 13 ^{+50}_{-2}$ MeV and a width of $294 \pm 25^{+46}_{-19}$ MeV}~\cite{COMPASS:2009xrl}.
Later in 2015, the COMPASS Collaboration observed $a_4$ in the $\pi^-p \to \eta^{(\prime)}\pi^-p$ process with {a} mass and width of $1900^{+80}_{-20}$ MeV/$c^2$ and $300^{+80}_{-100}$ MeV/$c^2${, respectively}~\cite{COMPASS:2014vkj}.
In 2018, COMPASS  performed {a most}  comprehensive resonance-model fit of $\pi^-\pi^-\pi^+$ states {based on their PWA} of a large data set of diffractive-dissociation events from the reaction $\pi^- + p \to \pi^-\pi^-\pi^+ +p_{\text{recoil}}$ with a 190 GeV/$c$ pion beam{.} {The} mass and width of $a_4(1970)$ {were reported as} $1935^{+11}_{-13}$ MeV and $333^{+16}_{-21}$ MeV, respectively~\cite{COMPASS:2018uzl}.
The most recent measurement of $a_4(1970)$ by  COMPASS indicates that the mass and width of $a_4(1970)$ are $1952.2 \pm 1.8 ^{+3}_{-3.5}$ MeV and $327 \pm 4 \pm 6$ MeV, respectively~\cite{Beckers2025}.

\par

\begin{figure}[htbp]
\centering
\includegraphics[scale=0.9]{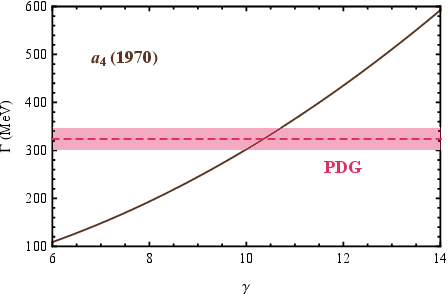}
\caption{{The $\gamma$ dependence of the total decay width of $a_4(1970)$, where the corresponding experimental data are shown for comparison with our theoretical calculation. The experimental values are taken from the PDG \cite{ParticleDataGroup:2024cfk}.} }
\label{a41970}
\end{figure}

The mass of $a_4(1970)$ {obtained in our calculation} is 1928 MeV, as shown in Table~\ref{mass}, which is slightly lower than that in {Refs.}~\cite{Steph:1985ff, Ebert:2009ub}, and is closest to the experimental value~\cite{ParticleDataGroup:2024cfk}.
The OZI-allowed two-body strong decay behavior of $a_4(1970)$ {is presented} in Table \ref{1F2F}. The total width of $a_4(1970)$ is 312 MeV according to our calculation, which {agrees} well with the experimental width {of} $324^{+15}_{-18}$ MeV~\cite{ParticleDataGroup:2024cfk}. The  {partial} decay widths of $\rho\omega$, $\pi\rho$, $\pi b_1$, {and} $\pi f_2$ are {found to be} 106 MeV, 68.5 MeV, 55.3  MeV and 33.7 MeV, respectively.  
It is worth {noting} that the branching ratio {$\Gamma(\rho \pi) /\Gamma(f_2\pi)$ is calculated to be $2.0$}, which is consistent with the experimental {value of} 1.7$^{+0.9}_{-0.8}$~\cite{Chung:2002pu, COMPASS:2018uzl}. In addition, the ratio $\Gamma(\eta^\prime \pi)  / \Gamma(\eta \pi )$ is \sout{abtained} {obtained} as $0.2$, in agreement with the experimental {value of} 0.23$\pm0.07$~\cite{GAMS:1995rpx, COMPASS:2014vkj}. 
{ We also consider the $\gamma$ dependence of the total decay width of $a_4(1970)$ in the range of 6$\sim$14, as shown in Fig. \ref{a41970}. The corresponding experimental data for comparison with our theoretical calculation are also presented. We find that when $\gamma$ is between 10 and 10.75, our theoretical calculation matches the experimental width of $a_4(1970)$, as depicted in Fig.~\ref{a41970}. As mentioned in Ref.~\cite{Pang:2025esm}, the ratios of the two-body strong decay channels are independent of the value of $\gamma$.}


\begin{figure}[htbp]
\centering
\includegraphics[scale=0.9]{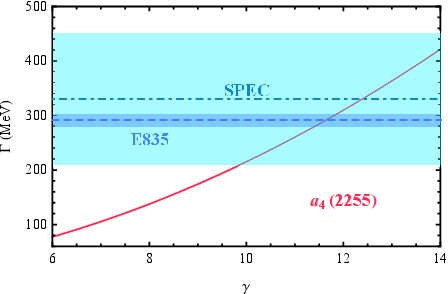}
\caption{{The $\gamma$ dependence of the total decay width of $a_4(2255)$, where the corresponding experimental data are shown for comparison with our theoretical calculation. The experimental values are taken from SPEC \cite{Anisovich:2001pn} and E835 \cite{Uman:2006xb}.}} 
\label{a42255}
\end{figure}

\begin{figure}[htbp]
\centering
\includegraphics[scale=0.9]{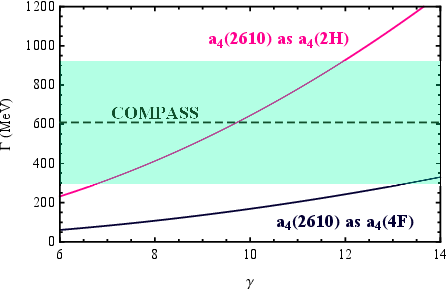}
\caption{{The $\gamma$ dependence of the total decay width of $a_4(2610)$ as $a_4(2H)$ and $a_4(2610)$ as $a_4(4F)$, where the corresponding experimental data are shown for comparison with our theoretical calculation. The experimental values are taken from COMPASS \cite{Beckers2025}.}    }
\label{a42610}
\end{figure}

\begin{table*}[htb] 
 \renewcommand{\arraystretch}{1.5}
\centering
\caption{The mass spectra of $a_4$ states. The unit is MeV. \label{mass}}
\[\begin{array}{ccccccccccccccc} 
\hline
\hline
&  & n^{2s+1}L_J &\text{State}  & \text{This work}& \text{GI}~$\cite{Steph:1985ff}$&\text{Ebert}~$\cite{Ebert:2009ub}$&\text{Exp. }$\cite{ParticleDataGroup:2024cfk}$
\\\midrule[0.7pt]
&  & 1^3F_4  &a_4(1970)    &1928  &2008  &2018 & 1967\pm16~$\cite{ParticleDataGroup:2024cfk} $    \\
&  &  2^3F_4  &a_4(2255)    &2243&2408 & 2284   &  2255\pm 40~$\cite{Uman:2006xb}$, 2237\pm55 ~$\cite{Anisovich:2001pn}$ \\
&  &  3^3F_4  & a_4(3F)   &2466 &2774& ... &  ...                \\
&  &  4^3F_4  & a_4(4F)     &2640  &3114 &...&2608\pm9^{+5}_{-38}~$\cite{Beckers2025}$ ?  \\
&  &  1^3H_4 & a_4(1H)     & 2405 &2643 &2234&  ...        \\
&  &  2^3H_4 &  a_4(2H)    &  2589&2970 &...&2608\pm9^{+5}_{-38}~$\cite{Beckers2025}$ ?             \\
 \hline
 \hline
\end{array}\] 
\end{table*}

The $a_4(2255)$ {is} listed in the “Further States” in PDG. It was first observed in 2001 with {a} mass and width {of} $2255\pm40$ MeV and $330^{+110}_{-50}$ MeV, respectively~\cite{Anisovich:2001pn}.  
Uman \textit{et al.} {reported} a strong resonance, $a_4(2237)$, decaying into $\eta\pi$, with $M/\Gamma=(2237\pm5)/(291\pm15)$ MeV~\cite{Uman:2006xb}. {The} $a_4(2237)$ {resonance} is {the} $a_4(2255)$.
The mass of $a_4(2255)$ we calculated by the MGI model is 2243 MeV, which is in better agreement with the experimental  {values of} 2237 MeV~\cite{Uman:2006xb} and 2255 MeV~\cite{Anisovich:2001pn}.   
{The} $a_4(2255)$ is considered {to be} a strong candidate for the $2F$ state and the OZI-allowed two-body strong decay width is {calculated to be} 222 MeV, which is comparable {to} the experimental values~\cite{Anisovich:2001pn,Uman:2006xb}. The decay channels $\pi f_2(1640)$, $\rho\omega$, $\pi b_1(2P)$, {and} $\pi \rho(1450)$ make significant contributions to the total width of the $a_4(2255)$. {The channels} $\rho a_2$, $\pi \rho_3$, $\pi \eta(1295)$ and $\rho h_1$ have branching ratios of $7.4\%$, $5.6\%$, $4.7\%$ and $3.4\%${, respectively}. 
It is worth noting that the $\eta\pi$ channel, with a branching fraction of $0.5\%$, is one of the decay modes {through} which the $a_4(2255)$ resonance was discovered experimentally~\cite{Anisovich:2001pn,Uman:2006xb}.
More details can been seen in Table \ref{1F2F}.
{\ In Fig.~\ref{a42255}, the $\gamma$ dependence of the total decay width of $a_4(2255)$ within the range of $\gamma = 6 \sim 14$ is presented. Our calculation is consistent with the experimental widths from SPEC \cite{Anisovich:2001pn} when $\gamma$ is between 10 and 14, and with E835 \cite{Uman:2006xb}  when $\gamma$ is between 11.5 and 12.
}
Our results strongly favor interpreting $a_4(2255)$ as the $2^3F_4$ state, i.e., the first radial excitation of $a_4(1970)$, thereby reinforcing the reliability of the MGI+QPC framework in describing the $a_4$ meson family and 
providing a basis for the discussion of even higher excitations such as $a_4(2610)$.


\begin{table}[htbp]
\renewcommand{\arraystretch}{1.4}
\centering
\caption{{Total} and partial decay widths of the  $a_4(1F)$ and $a_4(2F)$ states. “Br” denotes the branching ratio of the decay channel{. The} unit of width is MeV. {Channels} with widths {smaller} than 1 MeV are omitted.\label{1F2F}}

\[\begin{array}{cccccc}
\hline
\hline
 \text{Channel} & \text{Width} & \text{Br} & \text{Channel} & \text{Width} & \text{Br} \\
 \midrule[0.7pt]
 a_4\text{(1970)(1F)}   & 312   & 1    & a_4\text{(2255)(2F)}    & 222 & 1 \\
 \rho \omega            & 106   & 34.1 & \pi f_2\text{(1640)}    & 29.2 & 13.2 \\
 \pi \rho               & 68.5  & 22   & \rho \omega             & 27.4 & 12.4 \\
 \pi b_1                & 55.3  & 17.7 & \pi b_1\text{(2P)}      & 26 & 11.7 \\
 \pi f_2                & 33.7  & 10.8 & \text{$\pi \rho $(1450)}& 25.8 & 11.7 \\
 \pi \eta               & 12.3  & 3.94 & \rho a_2                & 16.4 & 7.39 \\
\pi \rho_3 & 6.51  & 2.09 & \pi \rho_3 & 12.5 & 5.64 \\
\pi \eta (1295)         & 5.73  & 1.84 & \pi \eta (1295)         & 10.4 & 4.7 \\
 \pi f_1                & 5.47  & 1.76 & \rho h_1                & 7.59 & 3.42 \\
 \pi \eta _2            & 3.72  & 1.19 & \pi b_1                 & 7.54 & 3.4 \\
  K^*K^* & 2.89 & 0.929 & \rho a_1 & 5.85 & 2.64 \\
 \text{$\pi \eta $'} & 2.45 & 0.785 & \pi \eta _2 & 5.17 & 2.34 \\
 \text{$\pi \rho $(1450)} & 2.41 & 0.774 & \pi \rho _2\text{(1D)} & 5.1 & 2.3 \\
 \pi \rho _2\text{(1D)} & 1.34 & 0.431 & \omega b_1 & 4.84 & 2.19 \\
 \text{...} & \text{...} & \text{...} & \pi f_2 & 4.52 & 2.04 \\
 \text{...} & \text{...} & \text{...} & \pi f_1\text{(2P)} & 4.12 & 1.86 \\
 \text{...} & \text{...} & \text{...} & \pi \rho _3\text{(1990)} & 4 & 1.81 \\
 \text{...} & \text{...} & \text{...} & \text{$\eta \pi $(1300)} & 2.72 & 1.23 \\
 \text{...} & \text{...} & \text{...} & \text{$\rho \omega $(1420)} & 2.55 & 1.15 \\
 \text{...} & \text{...} & \text{...} & \text{$\rho \pi $(1300)} & 2.28 & 1.03 \\
 \text{...} & \text{...} & \text{...} & \text{$\pi \eta $(1760)} & 2.23 & 1.01 \\
 \text{...} & \text{...} & \text{...} & \eta a_2 & 2.07 & 0.936 \\
 \text{...} & \text{...} & \text{...} & \pi f_3\text{(1F)} & 1.53 & 0.692 \\
 \text{...} & \text{...} & \text{...} & \pi \rho  & 1.34 & 0.605 \\
 \text{...} & \text{...} & \text{...} & \pi f_1 & 1.26 & 0.569 \\
 \text{...} & \text{...} & \text{...} & \pi \eta  & 1.23 & 0.553 \\
 \text{...} & \text{...} & \text{...} & \pi f_4 & 1.13 & 0.51 \\
 \text{...} & \text{...} & \text{...} & \pi b_3& 1.13 & 0.51 \\
 \hline\hline
\end{array}\]
\end{table}

\subsection{Study of the new state $a_4(2610)$}

Recently, COMPASS  observed a  {broad} shoulder at {a} mass above {that of} the $a_4(1970)$ in {the} $KK$ final state{. Their} fit indicated that, {on top of the non-resonant background}, an additional $a_4$ state  {was} needed to accurately describe the {spectrum} of the $J^{PC}=4^{++}$ wave~\cite{Beckers2025}. 
The corresponding mass and width were determined to be $2608 \pm 9^{+5}_{-38}$ MeV and $609 \pm 22^{+35}_{-311}$ Mev, respectively~\cite{Beckers2025}.
\par
For this newly observed $a_4(2610)$, we test the possibility of  $a_4(2610)$ being the candidate {for} $a_4(4F)$ or $a_4(2H)$ {by analyzing its} mass spectrum and two-body strong decay {properties}.
\par
{According to} the analysis of {the} mass spectrum {(Table~\ref{mass})} by the MGI model, {the} theoretical mass of the $a_4(2H)$ state, $2589~\text{MeV}$, {is in closer agreement with} the experimental value {of} $2608 \pm 9^{+5}_{-38}~\text{MeV}$  than the predicted mass of the $a_4(4F)$ {state}, $2640~\text{MeV}$.
 \par
Subsequently, we {investigate} the two-body strong decay behaviors based on {that the $a_4(2610)$ is the candidate {for} $a_4(4F)$ or $a_4(2H)$,}  as shown in Table \ref{2H4F}.
When we assign $a_4(2610)$ as a candidate {for the} $a_4(4F)$ {state}, we obtain the total decay width of  174 MeV and the most important decay channels  $\rho a_2(1700)$, $\pi b_1(2240)$, $\rho a_2$, $\pi b_1(1960)$ and $\pi \rho_3(1990)$. {In addition, the} decay modes $\pi f_2(1950)$, $\pi b_1$, $\rho \omega_3$, $\pi \rho_3$ and $\pi \rho$ also make contribution to the total decay width. 

When we assign $a_4(2610)$ as a candidate {for the} $a_4(2H)$ {state},  the two-body strong decay width  is {found to be} 665 MeV. Besides,  $\pi \rho_4(2G)$, $\pi b_3(2244)$ and $\pi \eta_4(2G)$ are the mainly decay channels{, while}  $\pi b_3$, $\pi \rho_2(1D)$, $\pi \eta_2$ and $\pi f_3$  {serve as} important decay modes. {In addition,} $a_4(2610)$ {can also} decay into $\pi f_3(1F)$, $\pi \rho_4$, $\pi b_1$ and {other channels} exhibited in Table \ref{2H4F}.    
The predicted width of the $a_4(2H)$, $\Gamma^{\text{th}}_{a_4(2H)} = 665~\text{MeV}$,
is in much better agreement with the experimental value  {of}
$609 \pm 22^{+35}_{-311}~\text{MeV}$,
compared with the narrower theoretical width of the $a_4(4F)$, 
$\Gamma^{\text{th}}_{a_4(4F)} = 174~\text{MeV}$.

On the other hand, the branching ratio of the $KK$ decay channel may {provide} useful information, since the $a_4(2610)$ was discovered in the $KK$ final state.
In our calculation, $\text{Br}(KK)=0.005\%$ for $a_4(2H)$ and $\text{Br}(KK)=0.001\%$ for $a_4(4F)$, which are not listed in 
Table \ref{2H4F}. {This result} implies that {the $a_4(2610)$} is more likely {to be assigned as} the $a_4(2H)$ {state}.

{\ We show the $\gamma$ dependence of the total decay width for $a_4(2610)$ in the range of $\gamma = 6 \sim 14$ in Fig. \ref{a42610}.
From the figure, we find that for $\gamma$ between 7 and 12, if $a_4(2610)$ is interpreted as $a_4(2H)$, our theoretical calculation aligns with the experimental width of $a_4(2610)$ from COMPASS \cite{Beckers2025}. For $\gamma$ greater than 13.5 and with $a_4(2610)$ interpreted as $a_4(4F)$, the theoretical width overlaps with the experimental data. However, for $\gamma$ values above 13.5, the theoretical width of $a_4(1970)$ is larger than 500 MeV, which contradicts  the experimental value \cite{ParticleDataGroup:2024cfk}.
}

Based on a comprehensive analysis of the mass spectrum and  two-body strong decays {under the} $a_4(4F)$ and $a_4(2H)$ {assignments}, we suggest that the newly {observed} $a_4(2610)$ is more likely to be $a_4(2H)$ {state} and $a_4(4F)$ may be a narrow state with {a} width  {of} 170 MeV. However, due to the large uncertainties ($609 \pm 22^{+35}_{-311}$) {on} width, more data will be required in the future for further verification.

\subsection{Prediction of the $a_4(1H)$ and $a_4(3F)$}
The $a_4(1H)$ and  $a_4(3F)$ {states} have {not yet been} observed in experiments. {Here, we provide}  predictions for their mass and {OZI-allowed two-body} strong decay behaviors.
{According to the MGI model, the predicted masses are 2405 MeV for $a_4(1H)$ and 2466 MeV for $a_4(3F)$. } Employing the spatial wave functions of $a_4(1H)$ and $a_4(3F)$ obtained from the MGI model as input, we calculated the OZI-allowed two-body strong {decay widths using the $^3P_0$ model, with results summarized} in Table \ref{1H3F}.
\par
As for $a_4(1H)$, we predict that its width is 685 MeV. The $\pi b_3$, $\pi \rho_4(2230)$, and $\pi f_3(1F)$ with branching ratios  {of} $20.4\%$, $15.6\%$, and $10.3\%$ are the {mainly} decay modes. $\pi \rho_2(1D)$, $\pi \eta_2$, $\rho a_1$, and $\rho \omega$ {also make significant contributions} to the total decay width. 
More {detailed} decay channels and their corresponding branching ratios can be seen in Table \ref{1H3F}.    

As for $a_4(3F)$, the OZI-allowed two-body strong decay width is predicted  to be about 250 MeV.   $\pi b_1(1960)$, $\pi f_2(1950)$, $\rho\omega$, $\rho a_2$, and $\pi \rho_3$ are the most important decay final state of $a_4(1H)${. In addition,} $a_4(1H)$ {can also} decay into $\pi\rho(1900)$, $\pi \rho_3(1990)$, $\pi f_2(1640)$, and $\pi \eta(1295)$. More detailed decay information is listed in Table \ref{1H3F}. 


\begin{table}[htbp]
\renewcommand{\arraystretch}{1.2}
\vspace{-0.8cm}
\centering
\caption{{Total} and partial decay widths of the   $a_4(2H)$ and $a_4(4F)$ states. The unit of width is MeV. {Channels} with widths {smaller} than 1 MeV are omitted.\label{2H4F}}
\[\begin{array}{cccccc}
\hline\hline
 \text{Channel} & \text{Width} & \text{Br} & \text{Channel} & \text{Width} & \text{Br} \\
 \hline
 a_4\text{(2610) as (2H)}  & 665 & 1     & a_4\text{(2610) as (4F)} & 174 & 1 \\
 \pi \rho _4\text{(2G)}    & 89.9 & 13.5 & \rho a_2\text{(1700)} & 10.5 & 6.04 \\
 \pi b_3\text{(2244)}      & 81.6 & 12.3 & \pi b_1\text{(2240)} & 10.5 & 6.03 \\
 \pi \eta _4\text{(2G)}    & 55.4 & 8.32 & \rho a_2 & 9.33 & 5.35 \\
 \pi b_3    & 39.7 & 5.97 & \pi b_1\text{(1960)} & 8.62 & 4.95 \\
 \pi \rho _2\text{(1D)}    & 35.9 & 5.4  & \pi \rho _3\text{(1990)} & 8.37 & 4.8 \\
 \pi \eta _2               & 27.6 & 4.15 & \pi f_2\text{(1950)} & 6.78 & 3.89 \\
 \pi f_3     & 27.1 & 4.07 & \pi b_1 & 6.05 & 3.47 \\
 \pi f_3\text{(1F)}        & 23.4 & 3.51 & \rho \omega _3 & 5.6 & 3.21 \\
 \pi \rho _4\text{(2230)}  & 21.5 & 3.23 & \pi \rho _3 & 5.49 & 3.15 \\
 \pi b_1                   & 20.3 & 3.05 & \pi \rho  & 5.16 & 2.96 \\
 \pi \rho _2\text{(2D)}    & 14.8 & 2.23 & \pi f_2 & 4.78 & 2.74 \\
 \eta a_3   & 12.5 & 1.88 & \rho \omega  & 4.68 & 2.68 \\
 \rho a_2                  & 10.5 & 1.58 & \pi b_1(2P) & 4.5 & 2.58 \\
 \pi b_5\text{(1H)}        & 10.5 & 1.58 & \pi \eta (1760) & 4.4 & 2.52 \\
 \pi b_1\text{(1960)}      & 10.1 & 1.51 & \pi \rho (4S) & 4.07 & 2.34 \\
 \pi \rho _2\text{(3D)}    & 9.77 & 1.47 & \pi \rho _2\text{(2D)} & 3.99 & 2.29 \\
 \eta \pi _2\text{(1880)}  & 9.42 & 1.42 & \pi \rho _3\text{(2250)} & 3.92 & 2.25 \\
 \rho a_1\text{(1640)}     & 9.18 & 1.38 & \rho h_1 & 3.88 & 2.23 \\
 \pi \eta _2\text{(2D)}    & 9.07 & 1.36 & \pi f_3\text{(2300)} & 3.73 & 2.14 \\
 \rho \omega (1420)        & 8.46 & 1.27 & \pi f_2\text{(1640)} & 3.6 & 2.07 \\
 \pi \rho _3(1990)         & 7.74 & 1.16 & \omega \rho _3 & 3.39 & 1.95 \\
 \rho h_1                  & 7.5 & 1.13  & \pi \eta _2(2D) & 3.2 & 1.83 \\
 \pi \eta _2\text{(3D)}    & 7.2 & 1.08  & \rho \omega (1420)& 3.03 & 1.74 \\
 \pi \rho _3  & 7.2 & 1.08  & \pi \rho _4\text{(2G)} & 2.95 & 1.69 \\
 \eta \pi _2               & 7.17& 1.08  & \pi \rho _2\text{(3D)} & 2.94 & 1.69 \\
 \rho a_1                 & 6.23 & 0.936 & \pi \eta _2 & 2.57 & 1.47 \\
 \omega b_1               & 5.75 & 0.865 & \rho h_1(1595) & 2.5 & 1.43 \\
\omega \rho (1450)        & 5.47 & 0.822 &\pi \rho(1900) & 2.42 & 1.39 \\
 \pi f_3(2300)            & 4.72 & 0.71  & \omega b_1 & 2.26 & 1.3 \\
 \pi f_5(1H)              & 4.65 & 0.7   & \pi \eta _2\text{(3D)} & 2.03 & 1.16 \\
 h_1b_1                   & 4.53 & 0.681 & \pi \rho _2\text{(1D)} & 1.91 & 1.09 \\
 \eta a_3\text{(1875)}    & 4.42 & 0.665 & \rho a_1 & 1.82 & 1.04 \\
 \rho h_1\text{(1595)}    & 4.4  & 0.662 & \omega \rho (1450)& 1.67 & 0.958 \\
 \pi f_1                  & 4.29 & 0.645 &\rho \pi (1300) & 1.33 & 0.763 \\
 \pi f_2                  & 3.68 & 0.553 &\pi \eta (1295)  & 1.22 & 0.701 \\
 \rho \omega              & 3.26 & 0.49  & \pi \eta_4(2G) & 1.21 & 0.693 \\
 \pi f_1(3P)              & 2.86 & 0.43  & \pi f_3 & 1.2 & 0.691 \\
 \eta a_1                 & 2.83 & 0.425 &\pi \eta (4S)  & 1.19 & 0.684 \\
 \pi \rho _3(2250)        & 2.68 & 0.402 & \pi f_1(2P)  & 1.14 & 0.654 \\
 \pi \rho                 & 2.64 & 0.397 & \pi \rho _4  & 1.13 & 0.65 \\
 \pi f_2(1950)            & 2.55 & 0.384 & \eta a_2 & 1.07 & 0.614 \\
 \pi f_4(2300)            & 2.35 & 0.354 & \pi f_1 & 1.04 & 0.595 \\
 \omega b_1(2P)           & 2.13 & 0.32  & \dots & \dots & \dots \\
 \pi \rho (1900)          & 1.75 & 0.263 & \dots & \dots & \dots \\
 \rho \omega _2(1D)       & 1.73 & 0.26  & \dots & \dots & \dots \\
 \pi f_4           & 1.63 & 0.245 & \dots & \dots & \dots \\
 \eta a_2                 & 1.56 & 0.235 & \dots & \dots & \dots \\
 \omega \rho _2(1D)       & 1.54 & 0.231 & \dots & \dots & \dots \\
 \pi \eta _4        & 1.49 & 0.224 & \dots & \dots & \dots \\
 \rho a_2(1700)           & 1.43 & 0.215 & \dots & \dots & \dots \\
 \pi f_2(1640)            & 1.23 & 0.186 & \dots & \dots & \dots \\
 b_1\pi (1300)            & 1.18 & 0.177 & \dots & \dots & \dots \\
 \pi b_1(2240)            & 1.04 & 0.157 & \dots & \dots & \dots \\
 \hline\hline
\end{array}\]
\end{table}

\begin{table}[htbp]
\renewcommand{\arraystretch}{1.2}
\vspace{-0.8cm}
\caption{{Total} and partial decay widths of the predicted  $a_4(1H)$ and $a_4(3F)$  states. The unit of width is MeV. Channels with widths smaller than 1 MeV are omitted. \label{1H3F}}

\[\begin{array}{cccccc}
\hline\hline
 \text{Channel} & \text{Width} & \text{Br} & \text{Channel} & \text{Width} & \text{Br} \\
 \hline
 \text{Tot.(}a_4\text{(1H))} & 685 & 1  & \text{Tot.(}a_4\text{(3F))} & 249 & 1 \\
 \pi b_3       & 140 & 20.4      & \pi b_1\text{(1960)} & 30.2 & 12.2 \\
 \pi \rho _4    & 107 & 15.6      & \pi f_2\text{(1950)} & 21.3 & 8.59 \\
 \pi f_3(1F)          & 70.8 & 10.3     & \rho \omega  & 14.9 & 6.02 \\
 \pi \rho _2(1D)      & 67.2 & 9.81     & \rho a_2 & 14.2 & 5.71 \\
 \pi \eta _2          & 45.4 & 6.62     & \pi \rho _3 & 11.7 & 4.72 \\
 \rho a_1             & 34.9 & 5.09     & \pi \rho (1900)  & 10.4 & 4.17 \\
 \rho \omega          & 29.5 & 4.31     & \pi \rho _3\text{(1990)} & 9.44 & 3.8 \\
 \pi b_1              & 19.3 & 2.81     & \pi f_2\text{(1640)} & 8.37 & 3.37 \\
 \pi \rho _3  & 17 & 2.48       & \text{$\pi \eta $(1295)} & 8.29 & 3.34 \\
 \pi b_1(2P)          & 16.3 & 2.38     & \rho h_1 & 8.12 & 3.27 \\
 \rho h_1             & 15.9 & 2.31     & \pi b_1\text{(2P)} & 8.11 & 3.26 \\
 \eta \pi _2          & 15.1 & 2.21     & \pi b_1 & 7.3 & 2.94 \\
 \omega b_1           & 13.4 & 1.95     & \pi f_2 & 6.74 & 2.71 \\
 \rho a_2             & 12.7 & 1.85     & \text{$\pi \rho $(1450)} & 6.09 & 2.45 \\
 \pi \rho _2(2D)      & 10.5 & 1.54     & \text{$\rho \omega $(1420)} & 5.65 & 2.28 \\
 \pi \eta _2(2D)      & 7.64 & 1.11     & \omega b_1 & 5.3 & 2.13 \\
 \pi f_2              & 5.74 & 0.838    & \pi \rho _2\text{(1D)} & 5.02 & 2.02 \\
 \pi f_1              & 5.04 & 0.736    & \pi \eta _2 & 4.96 & 2 \\
 \pi f_3       & 4.65 & 0.678    & \pi f_3 & 4.72 & 1.9 \\
 \pi f_4      & 4.24 & 0.619    & \pi \rho _2\text{(2D)} & 4.43 & 1.78 \\
 \pi f_1(2P)          & 4.08 & 0.596    & \rho a_1 & 4.38 & 1.76 \\
 \eta a_1             & 3.71 & 0.541    & \text{$\pi \eta $(1760)} & 4.11 & 1.65 \\
 KK_2           & 3.4 & 0.497     & \pi \eta _2\text{(2D)} & 3.89 & 1.56 \\
 \rho \pi (1300)      & 3.37 & 0.492    & \pi \rho _3\text{(2250)} & 3.76 & 1.51 \\
 KK_1                 & 3.13 & 0.457    & \pi \rho  & 3.66 & 1.47 \\
 \eta a_2             & 2.8 & 0.409     & \pi b_3\text{(2244)} & 3.56 & 1.43 \\
 \pi f_2(1640)        & 2.76 & 0.403    & \rho \omega _3 & 2.9 & 1.17 \\
 \pi \rho (1450)      & 2.61 & 0.38     & \pi \rho _4\text{(2230)} & 2.63 & 1.06 \\
 \pi \eta (1295)      & 2.24 & 0.327    &\omega \rho (1450) & 2.35 & 0.948 \\
 \pi \rho _3(1990)    & 2.12 & 0.309    & \eta a_2 & 2.32 & 0.932 \\
 \pi \rho             & 1.58 & 0.231    & \pi f_1\text{(3P)} & 2.29 & 0.921 \\
 \eta \pi (1300)      & 1.34 & 0.196    & \rho h_1\text{(1595)} & 1.69 & 0.681 \\
 \eta a_1(1640)       & 1.22 & 0.178    & K^*K_2^* & 1.47 & 0.591 \\
\dots & \dots & \dots                   & \pi f_1\text{(2P)} & 1.42 & 0.572 \\
\dots & \dots & \dots                   & \pi f_1 & 1.38 & 0.553 \\
\dots & \dots & \dots                   & \pi f_4 & 1.37 & 0.55 \\
\dots & \dots & \dots                   & \text{$\eta \pi $(1300)} & 1.25 & 0.504 \\
 \hline\hline
\end{array}\]
\end{table}

\begin{figure}[htbp]
\centering
\includegraphics[scale=0.9]{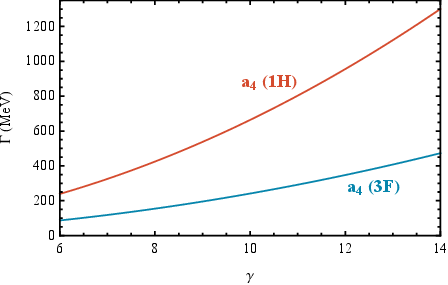}
\caption{{The $\gamma$ dependence of the total decay width of $a_4(1H)$ and $a_4(3F)$ }.}
\label{a41H3F}
\end{figure}

{We also present the $\gamma$ dependence of the total decay width of $a_4(1H)$ and  $a_4(3F)$ in the range of $\gamma = 6 \sim 14$ in Fig.~\ref{a41H3F}. We find that the predicted total width of $a_4(1H)$ and $a_4(3F)$ are in the ranges of 250$\sim$1300 MeV and 100$\sim$470 MeV, respectively.}

\section{CONCLUSION}\label{sec4}
In this work, we have systematically studied the mass spectra and  OZI-allowed two-body strong decay behaviors of {the} $a_4$ family, especially the newly observed $a_4(2610)$ state. 
\par
Our study indicates that for the already established ground state of the $a_4$ meson family, namely the $a_4(1970)$, the mass and decay properties predicted by the model {we used} are in very good agreement with experimental data.  
Likewise, the $a_4(2255)$ can be well interpreted as the first radial 
excitation of the $a_4(1970)$, with the theoretical predictions for both its mass and width being consistent with measurements. 
\par
For the newly observed $a_4(2610)$, we have examined its possible assignments as the $a_4(4F)$ and  $a_4(2H)$ states. {Both} the mass spectrum {and} the total width 
support {that} $a_4(2610)$ is more likely to be the $a_4(2H)$ state{, while} $a_4(4F)$ may {correspond to a relatively} narrow state with a width of 170 MeV.

In addition, we predict the masses and widths of the $a_4(3F)$ and $a_4(1H)$ states to be 
\begin{equation*}
\begin{cases} 
M(a_4(3F))= 2466~\text{MeV}\\
\Gamma(a_4(3F))= 250~\text{MeV}
\end{cases}
\begin{cases} 
M(a_4(1H))= 2405~\text{MeV}\\
\Gamma(a_4(1H))= 685~\text{MeV}
\end{cases}.
\end{equation*}
We also provide their dominant decay modes. These results may serve as valuable guidance for the experimental identification and {future searches} of the $a_4(3F)$, $a_4(4F)$, and $a_4(1H)$ states.

We look forward to upcoming experimental studies{, which will be crucial for} clarifying the nature of the newly observed light meson family members with $I^GJ^{PC}=1^{-}4^{++}$, as well as in validating or exploring these theoretical predictions {presented here}.

\begin{acknowledgments}
Y.-R. W. and C.-Q. P.  contributed equally to this work and should be considered co-first authors.
This work is supported by the National Natural Science Foundation of China under Grants  No.~12235018, No.~11975165, No.~11965016, and No.~12247101, and by the Natural Science Foundation of Qinghai Province under Grant No. 2022-ZJ-939Q, the Fundamental Research Funds for the Central Universities (Grant No. lzujbky-2024-jdzx06).
\end{acknowledgments}
\bibliographystyle{apsrev4-1}
\bibliography{hepref}
\end{document}